%% file: root.tex
\documentclass[letterpaper, 10 pt, conference]{ieeeconf}  % Comment this line out if you need a4paper

\IEEEoverridecommandlockouts                              % This command is only needed if 
\input{Packages}
\input{CustomCommands}
\title{\LARGE \bf
Adaptive Meshing for CPA Lyapunov Function Synthesis*
}

\author{Amy K. Strong$^{1}$, I. Samuel Akinwande$^{2}$, and Leila J. Bridgeman$^{1}$% <-this % stops a space
\thanks{*This work was supported by NSF Grant CMMI-2303157. Any opinions, findings, and conclusions or recommendations expressed in this material are those of the authors and do not necessarily reflect the views of the NSF.}% <-this % stops a space
\thanks{$^{1}$Thomas Lord Department of Mechanical Engineering and Materials Science, Duke University 
        {\tt\small amy.k.strong@duke.edu}}%
\thanks{$^{2}$Department of Aeronautics and Astronautics, Stanford University}%
}

\begin{document}

\maketitle
\thispagestyle{empty}
\pagestyle{empty}

%%%%%%%%%%%%%%%%%%%%%%%%%%%%%%%%%%%%%%%%%%%%%%%%%%%%%%%%%%%%%%%%%%%%%%%%%%%%%%%%
\begin{abstract}
\Ac{cpa} Lyapunov function synthesis is one method to perform Lyapunov stability analysis for nonlinear systems. This method first generates a mesh over the region of interest in the system's state space and then solves a \ac{lp}, which enforces constraints on each vertex of the mesh, to synthesize a Lyapunov function. Finer meshes broaden the class of Lyapunov function candidates, but \ac{cpa} function synthesis is more computationally expensive for finer meshes -- particularly so in higher dimensional systems. This paper explores methods to mesh the region of interest more efficiently so that a Lyapunov function can be synthesized using less computational effort. Three methods are explored --  adaptive meshing, meshing using knowledge of the system model, and a combination of the two. Numerical examples for two and three dimensional nonlinear dynamical systems are used to compare the efficacy of the three methods.
\end{abstract}

%%%%%%%%%%%%%%%%%%%%%%%%%%%%%%%%%%%%%%%%%%%%%%%%%%%%%%%%%%%%%%%%%%%%%%%%%%%%%%%%
\section{INTRODUCTION}

Lyapunov functions are a fundamental tool for stability analysis of nonlinear dynamical systems. A scalar, positive definite Lyapunov function, $V$, represents the energy of the system \cite{khalil2002}. The system is stable in the \ac{roa} where energy decreases over time, i.e. the \ac{pdi} $\dot{V}(\x) \leq 0$ holds for all $\x$ in the \ac{roa}. 

Synthesizing Lyapunov functions for nonlinear systems is a non-trivial problem. %Current research focuses on developing systematic ways to synthesize Lyapunov functions. 
Typically, a function form (\acf{cpa} \cite{giesl2012construction, giesl2014revised}, polynomial \cite{papachristodoulou2002construction}, neural network \cite{abate2020formal}, etc.) is used to represent the Lyapunov function. The function parameters are learned while encouraging or enforcing that the function adheres to Lyapunov function conditions. % for the particular system. %synthesizing the function for a particular system. %, develops constraints to enforce certain characteristics of the function, and then synthesizes the function via optimization methods. 
\Ac{cpa} Lyapunov function synthesis, in particular, applies to a broad class of systems, while maintaining true guarantees of stability (i.e., adhering to the Lyapunov function conditions). However, the synthesis process has high computational cost. % this benefit often comes at the cost of computation. 

A \Ac{cpa} Lyapunov function is defined on a mesh over a region of the system's state space, where a \acf{lp} assigns the function value at each vertex, uniquely defining it as affine on each simplex \cite{giesl2012construction,giesl2014revised}. The \ac{lp} imposes constraints at the mesh's vertices to ensure the \ac{cpa} function adheres to Lyapunov function conditions over all simplices. Thus, the number of simplices of the mesh directly affects the computational expense of the \ac{lp}. This paper explores computationally efficient methods of \ac{cpa} Lyapunov function synthesis via mesh generation and refinement strategies.

%In \ac{cpa} Lyapunov function synthesis, a mesh composed of simplices is created over a region of a system's state space. A \ac{cpa} Lyapunov function is then determined by assigning function values to each vertex in the mesh via an \ac{lp} \cite{giesl2012construction,giesl2014revised}. This \ac{lp} imposes constraints on the function across each simplex of the mesh to ensure that the synthesized function adheres to the desired Lyapunov characteristics. While constraints imposed on the mesh ensure the \ac{cpa} function is a true Lyapunov function,  this creates a computationally expensive \ac{lp} -- particularly when a finer mesh is needed.
%An advantage of \ac{cpa} Lyapunov function synthesis is that given an exponentially stable and Lipschitz continuous dynamical system, there is a guarantee that a \ac{cpa} Lyapunov function can be found so long the mesh is refined enough, i.e. the mesh has a high enough number of simplices \cite{giesl2014revised}. -- taking this out since it's for a specific mesh they use so maybe not directly applicable
%However, meshing a space is costly -- in high dimensions, prohibitively so. Moreover, the \ac{lp} designed requires multiple constraints on each simplex of a mesh -- creating a computationally expensive program. 

Current methods of \ac{cpa} function synthesis %use meshes of the state space -- creating a uniform grid mesh of the space use a 
use a simple mesh, where a fan of simplices is centered at the system's equilibrium and a uniform grid mesh is used elsewhere \cite{giesl2012existence, giesl2014revised}. We explore alternative methods of meshing for \ac{cpa} Lyapunov function synthesis, with the goal of synthesizing a valid Lyapunov function using fewer simplices compared to a naive grid mesh. We are motivated by the success of anisotropic meshes and adaptive mesh refinement in numerical analysis methods used to solve \acp{ode} and \acp{pde} \cite{alauzet2016decade,bouchard2012anisotropic,mesri2016optimal}. It has been observed that targeting difficult regions with smaller simplices while maintaining a coarser mesh elsewhere leads to more accurate solutions with fewer simplices \cite{alauzet2016decade}. However, formulating anisotropic mesh is an active area of research. A priori analysis of errors or the Hessian are often used to determine initial vertex placement of the mesh \cite{alauzet2016decade}. Adaptive mesh refinement is also popular. For example, a posteriori error estimates are used for mesh refinement strategies \cite{yano2012optimization} or reinforcement learning has been used to target mesh refinement \cite{freymuth2023swarm}. However, to our knowledge, mesh refinement and generation strategies have not yet been explored for \ac{cpa} Lyapunov function synthesis. %to the solution of Lyapunov inequalities%, as we have here.

%This aim is motivated by anisotropic mesh adaptation in \ac{cfd} \cite{alauzet2016decade} and \acp{fem} for structural problems [cite] to solve \acp{pde}. In adaptive solvers, a priori analysis of the relevant \ac{pde} is often used to determine initial vertex placement of the mesh, while a posteriori error analysis dictates mesh refinement to target difficult regions. The overall result is more accurate solutions with fewer simplices when compared to a uniform refinement [cite]. However, care must be taken in mesh refinement to ensure that regularity of the mesh is maintained [cite].
%
\begin{comment}
\begin{itemize}
    \item For odes/fem, mesh refinement is often considered in terms of h- and p-, where p is the polynomial used (the extent of the taylor series' expansion) and h- is the size of the simplices. The thought is that you can get a solution if p -> inf or h -> 0. Here, we have a fixed p- and are considering h-. However, we do use the fixed p- when considering vertex placement in method 2. 
    \item 
\end{itemize}
\end{comment}
%To do: add somewhere in here that meshing for FEM is often only ever going to be considered for 3D, whereas we need to consider potentially more dimensions. This motivates our choice for methods that can more easily adapt to high dimensions

This paper considers three methods of mesh formulation to efficiently synthesize a \ac{cpa} Lyapunov function for a nonlinear system. Method 1 introduces slack variables into the original \ac{cpa} Lyapunov \ac{lp} to guide an adaptive meshing procedure. Method 2 considers the \ac{pdi}, $\dot{V}(\x) \leq 0$, a priori; we estimate changes in the Hessian of the \ac{pdi} and leverage that information for vertex placement of the mesh. Finally, method 3 combines a priori and a posteriori mesh refinement; the mesh developed in method 2 is iteratively adapted using method 1. Numerical examples explore the efficacy of these methods compared to a naive grid meshing when synthesizing Lyapunov functions for two and three dimensional nonlinear systems. 

\subsubsection*{Notation}
The interior, boundary, and closure of the set $\Omega \subset \IR ^n$ are denoted as $\Omega^o,$ $ \delta \Omega,$ and $\overline{\Omega},$ respectively. The symbol $\mathfrak{R}^n$ denotes the set of all compact subsets $\Omega \subset \IR ^n$ satisfying i) $\Omega^o$ is connected and contains the origin and ii) $\Omega = \overline{\Omega^o}$. Scalars and vectors are denoted as $x$ and $\x,$ respectively. The notation $\mathbb{Z}_{a}^b$ denotes the set of integers between $a$ and $b$ inclusive. The $p$-norm of the vector $\x \vecdim{n}$ is shown as $||\x||_p,$ where $p \in \mathbb{Z}_1^{\infty}.$ 
By $f \in \mathcal{C}^k(\Omega)$, it is denoted that a real valued function, $f$, is $k$-times continuously differentiable over its domain $\Omega$. Let $1_n$ denote a vector of ones in $\IR ^n.$ %The Dini derivative of a function, $f$, is denoted as $D^+f(\x)$  \cite{giesl2014revised}.

%The right-hand (left-hand) upper Dini derivatives for some function, $f$  is defined as $D^{+}f(\y) \defeq \lim \sup_{k \rightarrow 0^{+}} \!\!\frac{f(\x {+} kg(\x)) {-} f(\x)}{k}\! \left(\!\lim \sup_{k \rightarrow 0^{-}} \!\!\frac{f(\x {+} kg(\x)) {-} f(\x)}{k}\!\right)\!\!,$ where $k \in \IR $ and $\dot{\x} = g(\x)$ \cite{giesl2014revised}.
\section{Preliminaries}
This paper explores meshing schemes for \ac{cpa} Lyapunov functions synthesis of a nonlinear dynamical system, defined below. The following sections detail mesh generation and prior work on \ac{cpa} Lyapunov function synthesis \cite{giesl2012construction,giesl2014revised} that will be used for this task.
\begin{definition}\label{def:dynSys}
Define the dynamical system,
\begin{equation}\label{eq:dynSys}
    \dot{\x} = f(\x),
\end{equation}
where $f{:}\IR^n{\rightarrow}\IR^{n}$ and  $f{\in}\mathcal{C}^2(\IR^n)$. Let \eqref{eq:dynSys} be locally exponentially stable in $\Omega{\in} \mathfrak{R}^n$ with equilibrium $\x =\mathbf{0}$.
\end{definition}
%, and outlines the problem statement. %To do? briefly talk about simple mesh generation?

\subsection{Mesh}
%\subsubsection{Basic Tools}
%The definitions and tools for mesh generation and refinement are described below.
%\begin{definition} (\textit{Affine independence} 
%\cite{giesl2014revised}):
%    A collection of $m$ vectors $\{\x_0, \x_1, \hdots , \x_m\} \subset \IR ^n$ is affinely independent if $\x_1-\x_0, \hdots, \x_m - \x_0$ are linearly independent.  
%\end{definition}
\begin{definition} (\textit{$n$ - simplex} \cite{giesl2014revised}):
    A simplex, $\sigma$, is defined as the convex hull of $n{+}1$ affinely independent vectors in $\mathbb{R}^n$, %, $co\{\x_j\}_{j=0}^n$, 
    where each vector, $\x_{j} \in \IR ^n$ ($j\in\mathbb{Z}_0^n$), is a vertex. A \emph{face} is the convex hull of $m \leq n$ affinely independent vectors in $\mathbb{R}^n$.
\end{definition}
\begin{definition} (\textit{Mesh} \cite{giesl2014revised}):
    Let  $\Tcal = \{\sigma_i\}_{i=1}^{m_{\Tcal}} \in \mathfrak{R}^n$ represent a union of $m_{\Tcal}$ simplexes, where the intersection of any two simplexes is a face or an empty set.
\end{definition}
Let $\{\x_{i,j}\}_{j=0}^n$ be $\sigma_i$'s vertices. The vertex $\x_{i,0}$ in $\sigma_i$ is arbitrary unless $0 \in \sigma_i,$ in which case $\x_{i,0} = 0$ \cite{giesl2014revised}. Let $\Tcal$ be a mesh of $\Omega \in \mathfrak{R}^n$. The vertices of the mesh are denoted as $\mathbb{E}_{\Tcal},$ $\Tcal_{0}$ denotes simplices containing the origin, and $\Tcal_{\Omega\setminus\{0\}}$ denotes simplices not containing the origin. %Let $n_\Tcal$ denote the number of vertices in a mesh.

Bounding sets are later used for model-informed meshing.

\begin{definition}(Bounding Sets \cite{akinwande2025verifyingnonlinearneuralfeedback})
    A bounding set is defined as the tuple $\langle n,P,L,U\rangle$, where $n$ is a natural number, $P$ is a finite set of points in $\mathbb{R}^n$, and $L,U$ are functions from $P$ to $\mathbb{R}$, defined such that $L(p) \leq U(p) \, \forall \, p \in P$.
\end{definition}

\subsection{CPA Lyapunov Function Synthesis}

A \ac{cpa} function is finitely represented by the function value at each vertex, i.e.  $\W {=} \{W_{\x}\}_{\x \in \mathbb{E_{\Tcal}}}{\subset}\IR.$ The gradient of a \ac{cpa} function across a mesh is calculated via \Cref{lem:lemma_gradW}.

\begin{lemma}\cite[Remark 9]{giesl2014revised}\label{lem:lemma_gradW}
    Consider the mesh $\Tcal = \{\sigma_i\}_{i=1}^{m_{\Tcal}}$ and set $\W = \{W_{\x}\}_{\x \in \mathbb{E_{\Tcal}}}\subset\IR,$ where $W(\x) = W_{\x}, \forall \x \in \mathbb{E}_{\Tcal}.$ For $\sigma_i$, let $\X_i \vecdim{n \times n}$ be a matrix with $\x_{i,j} - \x_{i,0}$ as its $j$-th row and $\bar{W}_i\vecdim{n}$ be a vector with $W_{\x_{i,j}} - W_{\x_{i,0}},$ as its $j$-th element. The unique \ac{cpa} interpolation of $\W$ for $\x {\in }\sigma_i$ is $W(\x) = \x^\top \X_i^{-1}\bar{W}_i.$ %where $b_i$ is a bias term. 
\end{lemma}

In \ac{cpa} Lyapunov function synthesis, a mesh is created over a region of the state space, and the Lyapunov function conditions are enforced at the mesh vertices.%, as seen below in \Cref{thm:lyap}. 

\begin{theorem} \cite[Theorem 1]{giesl2014revised} \label{thm:lyap}
Consider the dynamical system in \Cref{def:dynSys}. %the dynamical system,
%\begin{equation}\label{eq:dynSys}
 %   \dot{\x} = f(\x),
%\end{equation}
%in $\Omega \red{\in} \mathfrak{R}^n$, where $f:\map{n}{n}$ and  $f\in\mathcal{C}^2(\Omega)$. 
Define a  mesh $\Tcal = \{\sigma_i\}_{i=1}^{m_{\Tcal}}$ of $\Omega$, $\mathbf{L} =\{l_i\}_{i=1}^{m_\Tcal} \subset \IR^n$, and define the \ac{cpa} function $V = \{V_\x\}_{\x\in\mathbb{E}_{\Tcal}}.$ Consider
	\begin{subequations}\label{eq:originalOpt}
		\begin{flalign}\label{eq:pos}
			&V_\x \geq \norm{\x}_2, \quad \forall \x \in \mathbb{E}_\Tcal 
		\end{flalign}
		\begin{flalign}\label{eq:grad}
			&\norm{\nabla V_i}_1 \leq l_i,  \quad i \in \mathbb{Z}_1^{m_\Tcal} 
		\end{flalign}
		\begin{flalign}\label{eq:decrease}
			&\nabla V_i^\top f(\x_{i,j}) + \frac{1}{2}c_{i,j}\beta_i 1_n^\top l_i  \leq  -\norm{\x_{i,j}}_2  \\  &\quad\quad\quad\quad\quad\quad\quad\quad\quad\forall i \in \mathbb{Z}_1^{m_\Tcal}, j \in \mathbb{Z}_0^n, \nonumber
		\end{flalign}
	\end{subequations}
    where $\nabla V_i$ is the gradient of V across simplex $\sigma_i$ (\Cref{lem:lemma_gradW}),
    \begin{flalign}
        \beta_i &\geq \max_{p,q,r\in\mathbb{Z}_0^n, \x \in\sigma_i} \absVal{\frac{\partial ^2f^{(q)}}{\partial x_r\partial x_s}}, \label{eq:beta} \\
        c_{i,j} &= \begin{cases} \label{eq:c}
            &\hspace{-3mm} n \max_{k \in \mathbb{Z}_0^n} ||\Delta \x^i_{j,k}||_2^2,  \quad \forall \sigma_i \in \Tcal_{\Omega\setminus \{0\}} \\
            &\hspace{-3mm} n||\Delta \x^i_{j,0}||_2(\max_{k \in \mathbb{Z}_1^n}||\Delta \x^i_{0,k}||_2 {+} ||\Delta \x^i_{j,0}||_2), \\ & \hspace{45mm} \forall \sigma_i \in\Tcal_0,
        \end{cases}
    \end{flalign}
    and $\Delta \x^i_{j,k} {=} \x_{i,j} {-} \x_{i,k}.$ If $V$ adheres to \eqref{eq:originalOpt}, then $V$ is a Lyapunov function for \eqref{eq:dynSys} in $\Omega$, and \eqref{eq:dynSys} is exponentially stable in $\Omega$.
\end{theorem}

In \Cref{thm:lyap}, \eqref{eq:pos} enforces positive definiteness, \eqref{eq:grad} bounds the gradient of the Lyapunov function, and \eqref{eq:decrease} enforces the Lyapunov decrease condition for exponential stability. In Constraint \eqref{eq:decrease}, the term $\frac{1}{2}c_{i,j}\beta_i1^\top l_i$ accounts for the dynamics of the system across a simplex and ensures the decrease condition holds across each simplex \cite{giesl2014revised}. % despite only being enforced at the vertices of the mesh. 
It is found by applying a second order Taylor series expansion \cite[Theorem 14.20]{fitzpatrick2009advanced} to the system dynamics and, thus, uses a bound on the Hessian \eqref{eq:beta} across each simplex \cite{giesl2014revised}. The term $c_{i,j}$ results from applying the expansion about any $\x\in\sigma_i$ if $\sigma_i\in\Tcal_{\Omega\setminus{0}}$  \cite{strong2024improved} or about $\x_{j,0}$ if $\sigma_i \in \Tcal_0$ \cite{giesl2014revised}.

%\begin{remark}
%    The term $c_{i,j}$ in \Cref{thm:lyap} results from applying a second order Taylor series expansion \cite[Theorem 14.20]{fitzpatrick2009advanced} to $f(\x)$ about any $\x\in\sigma_i$ \cite[Theorem 2]{strong2024improved} or about $\x_{j,0}$ \cite{giesl2014revised}.
%\end{remark}

%\red{Note that in \eqref{eq:c}, $c_{i,j}$ is defined using two different versions of a Taylor series expansion error bound.}

%For an exponentially stable, Lipschitz continuous system, \eqref{eq:originalOpt} will always have a solution if there are enough simplices, $m_\Tcal$, in the mesh (defined by \cite[Definition 13]{giesl2014revised}) \cite[Theorem 5]{giesl2014revised}.

\section{Motivation}

Current \ac{cpa} Lyapunov function literature typically uses a simple grid mesh with a fan structure near the origin \cite{giesl2014revised}. However, the mesh affects \ac{cpa} Lyapunov function synthesis. In~\eqref{eq:decrease}, $\frac{1}{2}c_{i,j}\beta_i 1_n^\top l_i$ depends on both the diameter of the simplex (via $c_{i,j}$) and the Hessian of $f$ across the simplex (via $\beta_i$). If a simplex with a large diameter overlays a region where $\beta_i$ is large, then~\eqref{eq:decrease} enforces a very strict decrease condition. If a simplex contains an evolving Hessian, the decrease condition may be stricter than needed in some areas. These issues may result in an infeasible problem given a mesh.

Theoretically, the effect of the mesh is not a pressing issue. A \ac{cpa} Lyapunov function can always be found for an exponentially stable, Lipschitz continuous system if the mesh has enough simplices \cite[Theorem 5]{giesl2014revised}. Thus, current \ac{cpa} Lyapunov function literature uniformly refines the grid mesh until a Lyapunov function is found. %However, this strategy can result in fine grid meshes with a large number of simplices being used, making \ac{cpa} Lyapunov function synthesis computationally expensive or even infeasible.  
However, \Cref{thm:lyap} requires $(2n{+}1)m_\Tcal + v_\Tcal$ constraints, where $n$ is system dimension, $m_\Tcal$ is number of mesh simplices, and $v_\Tcal$ is number of mesh vertices. For systems with large \acp{roa} or high dimensions, meshes become a computational bottleneck.
%that the mesh structure affects the Lyapunov function conditions via $\beta_i$ and $c_{i,j}.$ This often results in a very fine grid meshes being used -- resulting in a large number of simplices. This can make \ac{cpa} Lyapunov function synthesis computationally expensive or even infeasible, as in \Cref{thm:lyap}, each simplex requires $3n+2$ constraints. }
This paper thus explores alternate meshing methods to achieve \ac{cpa} Lyapunov function synthesis with smaller number of simplices than a naive grid mesh method.

\begin{comment}
Our objective is summarized as:
\begin{statement}\label{state:objective}
    Let
    \begin{equation}\label{eq:dynSys}
        \dot{\x} = f(\x),
    \end{equation}
    where $\x\vecdim{n}$, be Lipschitz continuous and exponentially stable in $\Omega \red{\in} \mathfrak{R}^n$ with an equilibrium point at the origin. Define a grid triangulation $\Tcal_G = \{\sigma_i\}_{i=1}^{m_{\Tcal_G}}$ of $\Omega$. 
    The objective of this paper is to synthesize a valid \ac{cpa} Lyapunov function, $V = \{V_\x\}_{\x\in\mathbb{E}_{\Tcal}}$ over $\Tcal = \{\sigma_i\}_{i=1}^{m_{\Tcal}}$ (i.e. $V$ adheres to \eqref{eq:originalOpt}) \red{using as few simplices as possible.}
    than required to synthesize the valid Lyapunov function $V_G = \{V_\x\}_{\x\in\mathbb{E}_{\Tcal_G}}$ over the triangulation $\Tcal_G$, i.e. $m_{\Tcal} <m_{\Tcal_G}.$
\end{statement}
\end{comment}

\section{MAIN RESULTS}

We explore two perspectives on mesh generation for \ac{cpa} Lyapunov function synthesis. First, we do not consider the system model a priori and instead use a relaxed optimization problem to guide online adaptation of the mesh. Method 1 iteratively refines an initial mesh -- targeting refinement to areas where the Lyapunov decrease condition is violated. Second, in method 2, we consider system dynamics -- specifically the evolution of the Hessian of system dynamics across the region -- when creating the mesh. Finally, we combine these two perspectives in method 3, using method 2 to generate an initial mesh, which is refined by method 1.

\subsection{Method 1: Online Adaptation}

Method 1 refines an initial mesh based on the regions where the Lyapunov decrease condition \eqref{eq:decrease} is most prohibitive to \ac{cpa} function synthesis. The initial mesh used is any sparse mesh for which \ac{cpa} function synthesis is not initially feasible. We introduce slack variables into \eqref{eq:originalOpt} to ease the decrease condition, \eqref{eq:decrease}. This creates a relaxed \ac{lp} that, when solved, indicates which vertices (and corresponding simplices) violate the decrease condition and to what degree. 
\Cref{cor:optSlack} reformulates \eqref{eq:originalOpt} with slack variables $\Upsilon = \{\upsilon_\x\}_{\x \in \mathbb{E}_\Tcal}$ assigned to each vertex of the mesh. 

\begin{corollary}\label{cor:optSlack}
	Consider the dynamical system in \Cref{def:dynSys}. Let $\Tcal {=} \{\sigma_i\}_{i=1}^{m_{\Tcal}}$ be a mesh of $\Omega$. Define $\mathbf{L} =\{l_i\}_{i=1}^{m_\Tcal}{ \subset} \IR^n$, the \ac{cpa} function $V {=} \{V_\x\}_{\x\in\mathbb{E}_{\Tcal}}$, and slack variables $\Upsilon {=} \{\upsilon_\x\}_{\x \in \mathbb{E}_\Tcal}$. Consider the optimization problem
	\begin{subequations}\label{eq:optSlack}
		\begin{equation*}
			\min_{\Upsilon, V, \mathbf{L}} \sum_{\x \in \mathbb{E}_\Tcal} \upsilon_{\x}
		\end{equation*}
		\begin{flalign}
			& V_\x \geq \norm{\x}_2, \quad \forall \x \in \mathbb{E}_\Tcal \label{eq:posNew} \\
			&\norm{\nabla V}_1 \leq l_i,  \quad \forall i \in \mathbb{Z}_1^{m_\Tcal} \label{eq:gradNew} \\
			& \nabla V_i^\top f(\x_{i,j}) + \frac{1}{2}c_{i,j}\beta_i 1_n^\top l_i +\norm{\x_{i,j}}_2 \leq  \upsilon_{\x_{i,j}} \label{eq:decreaseNew} \\  &\quad\quad\quad\quad\quad\quad\quad\quad\quad\quad\quad\quad\quad\forall i \in \mathbb{Z}_1^{m_\Tcal}, j \in \mathbb{Z}_0^n, \nonumber \\
			&\upsilon_{\x} \geq -\alpha, \quad \forall \x \in \mathbb{E}_\Tcal,
		\end{flalign}
	\end{subequations}
    where $\beta_i$ and $c_{i,j}$ are defined by \eqref{eq:beta} and \eqref{eq:c}, $\nabla V_i$ is the gradient of $V$ across simplex $\sigma_i$ (\Cref{lem:lemma_gradW}), and $\alpha>0$ is a pre-set constant. The \ac{lp} will always have a solution. If $\upsilon_\x \leq 0$ for all $\x \in \mathbb{E}_\Tcal$, the $V$ is a valid Lyapunov function.
\end{corollary}
\begin{proof}
    Since $\Omega$ is bounded, setting $V_\x=\norm{\x}_2$ is a valid solution of \eqref{eq:posNew} and induces finite, constant $\nabla V_i$ across each simplex, so setting $l_i=\norm{\nabla V_i}_1$ satisfies \eqref{eq:gradNew}. This also ensures that $c_{i,j}$ is always finite. Boundedness of $\Omega$ and continuity of $f$ then together ensure that all terms on the left-hand side of \eqref{eq:decreaseNew} are finite, making that equation feasible with finite $\upsilon_{\x_{i,j}}\geq -\alpha$.
\end{proof}

If the solution of \eqref{eq:optSlack} results in one or more positive slack variables, we iteratively refine the simplices with decrease condition violations until a viable Lyapunov function is found. We do so using \ac{leb} \cite{rivara1984algorithms, rivara19923, korotov2016longest}. To summarize \ac{leb}, a vertex is added on the longest edge of a targeted simplex -- bisecting the simplex into two new simplices. Neighboring simplices are also refined to ensure conformity of the mesh; for example, if the neighbor intersecting the target simplex's longest edge does not have a matching longest edge, %has a matching longest edge, then it is also bisected via the newly placed vertex. If the neighbor 
%has a longest edge not intersecting the target simplex, 
the neighbor is first bisected along its own longest edge before bisecting the resulting simplex that intersects with the target simplex again. Thus, \ac{leb} results in refinement propagating throughout the mesh. While refining neighboring simplices could be avoided by placing a new vertex in the interior of the target simplex, it is crucial to cut the longest edge of the simplex, as it affects the conservativeness of the \ac{lp} via $c_{i,j}$. 

The targeted refinement process is summarized in \Cref{alg:adaptTri}. The algorithm is initialized with some sparse mesh, $\Tcal$ and then iteratively solves \eqref{eq:optSlack}, determining the simplex with the largest violation of the decrease condition at each iteration (Lines 3-4). \ac{leb} then refines this simplex (Line 5).
\Cref{alg:adaptTri} iterates until a viable Lyapunov function is found, i.e., $\upsilon_\x \leq 0$ $\forall \x \in \mathbb{E}_\Tcal.$
\begin{algorithm}[h!]
	\caption{Adaptive mesh}
	\label{alg:adaptTri}
	\begin{algorithmic}[1]
		\Require $\Tcal$
		\While{$\exists \x\in\mathbb{E}_\Tcal$ s.t. $\upsilon_\x \not\leq 0$}
		\State $[V,\Upsilon,\mathbf{L}]=$ Solve \eqref{eq:optSlack}
		\State $\Sigma =\{\sum_{j=0}^n \upsilon_{i,j}\}_{i=1}^{m_\Tcal} $
		\State $\overline{\sigma} = \max_{i \in \mathbb{Z}_1^{m_\Tcal}} \Sigma$
		\State $\Tcal = $LEB $(\Tcal, \overline{\sigma})$
		\EndWhile
		\State \Return $V,$ $m_\Tcal$, $\Tcal$
	\end{algorithmic}
\end{algorithm}
 
\subsection{Method 2: Model Informed Mesh}

Method 2 considers system dynamics when creating an initial mesh. The error term in \eqref{eq:decrease} depends on the Hessian of the dynamical system across an individual simplex via $\beta_i$. 
%The intuition behind method 2 is to determine regions of $\Omega$ where $\beta_i$ will be high and reduce the size of the simplices in those regions accordingly.
Method 2 determines regions of $\Omega$ where $\beta_i$ evolves quickly and densely meshes those regions accordingly to ensure a tight bound. To do so, we leverage the OVERTPoly algorithm \cite{akinwande2025verifyingnonlinearneuralfeedback}. %to analyze individual dimensions of the nonlinear system. %OvertPoly \cite{akinwande2025verifyingnonlinearneuralfeedback} constructs polyhedral enclosures (i.e., piecewise linear upper and lower bounds) for nonlinear functions that are represented using rational compositions of univariate functions. %These bounds are used to compute forward reachable sets for nonlinear systems controlled by neural networks. An unintended consequence of 
The algorithm identifies these regions by decomposing the dynamics into univariate components and partitioning each univariate domain into intervals over which the convexity is invariant. It then reconstructs multivariate domains using the composition rules defined in~\cite{akinwande2025verifyingnonlinearneuralfeedback}. The result is an approximation of the evolution of the Hessian. In the context of this work, this corresponds to regions where $\beta_i$ changes. 
% The bound generation process %is that the OvertPoly algorithm 
% partitions the domain in a manner that approximates the evolution of a function's Hessian.

These operations are performed using bounding sets defined over finite point sets $P$ in each dimension. Although the algorithm was originally designed for bound computation, the resulting point sets $P$ implicitly partition the domain into regions over which the Hessian may vary. When $P$ is used to generate vertices, it enables the construction of a mesh informed by system dynamics. 
The resulting vertices are not guaranteed to contain the dynamical system's equilibrium, %(in \Cref{state:objective}, the origin), 
which must be a vertex to maintain feasibility of the \ac{lp}. Therefore, the axis $x_k = 0$ is added to the list of vertices for each dimension $k = 1, \hdots, n.$ %Essentially, an additional row of points is added where each dimension is $0$ -- including a point at the origin. 
To prevent irregular simplices, any point from the original list of vertices that is within $0.05$ of the origin is removed. Finally, a Delaunay triangulation is applied to the vertices to produce the mesh.

\subsection{Method 3: Combined Approach}

Method 3 explores online refinement (method 1) of a mesh that is formulated with the system dynamics in mind (method 2).
In short, method 3 is implemented by initializing \Cref{alg:adaptTri} with a mesh from method 2.

\section{NUMERICAL EXAMPLES}

We evaluated the capability of our methods for various two dimensional systems and a three dimensional system by comparing the number of simplices required for each method to successfully synthesize a \ac{cpa} Lyapunov function. A naive grid mesh is used as a baseline for comparison.
%In this section, \ac{cpa} Lyapunov functions are generated for various two and three dimensional nonlinear dynamical systems using different methods to mesh the region of attraction, $\Omega.$
%For each system, a table compares the performance of methods 1-3 to that of a standard grid mesh. 

\subsection{Two-Dimensional Nonlinear Systems}

We consider the following two-dimensional systems:
\begin{enumerate}
    \item System A: (Pendulum) The sinusoidal system is defined as
    \begin{equation}\label{eq:sysA}
	\dot{\x} = \begin{pmatrix}
            f_1(\x) \\
		     f_2(\x)
        \end{pmatrix}=
        \begin{pmatrix}
            x_2 \\
		     -\sin(x_1) -x_2,
        \end{pmatrix}
    \end{equation}
    over $\Omega= [-\frac{\pi}{2}, \frac{\pi}{2}]\times [-\frac{\pi}{2}, \frac{\pi}{2}].$ The nonlinearity of the system is isolated to $\dot{x}_2,$ so the Hessian of $f_1(\x)$ varies only based on the value of $\sin(x_1).$ Thus, method 2 does not bound $x_2$, but instead uses a uniform grid of $\frac{\pi}{6}$ for vertex points in this dimension.
    \item System B: The polynomial system is defined as
       \begin{flalign}\label{eq:sysB}
        \dot{\x} = \begin{pmatrix}
            f_1(\x) \\
		     f_2(\x)
        \end{pmatrix}=
            \begin{pmatrix}
    	         0.3 x_1^5-0.5 x_2^4 - 0.5x_1\\
    		   -0.5x_1^6-0.1x_2
    	\end{pmatrix}
    \end{flalign}
    over $\Omega= [-0.75, 0.75] \times [-0.75, 0.75]$. System 2 is nonlinear in both $x_1$ and $x_2$. The Hessian of $f_1(\x)$ depends on the values $6x_1^3$ and $-6x_2^2,$ while the Hessian of $f_2(\x)$ depends on $-15x_1^4.$ Each Hessian has no diagonal terms-- making it ideal for method 2.
    \item System C: The final system is defined as
    \begin{flalign}\label{eq:sysC}
        \dot{\x} = \begin{pmatrix}
            f_1(\x) \\
		     f_2(\x)
        \end{pmatrix} =
            \begin{pmatrix}
    	       0.5x_1^4\sin(x_2) + 0.3x_2 \\
    		   -0.5x_1-1.25x_2-x_2^3x_1 
    	\end{pmatrix}
    \end{flalign}
    over $\Omega= [-1, 1] \times [-1, 1]$. The two dimensions interact in the Hessian matrix. The Hessian of $f_1(\x)$ depends on the terms $-6x_1^2\sin(x_2)$, $2x_1^3\cos(x_2),$ $-\frac{1}{2}x_1^4\sin(x_2)$, and $2x_1^3\cos(x_2),$ while the Hessian of $f_2(\x)$ depends on $-3x_2^2$ and $-6x_1x_2.$ 
\end{enumerate}

\Cref{tab:results2D} shows the performance of %the meshes generated by 
methods 1-3 when used to synthesize \ac{cpa} Lyapunov functions for the above systems. In method 1, the mesh used to initialize \Cref{alg:adaptTri} is a grid with uniform spacing, whereas in method 3, \Cref{alg:adaptTri} is initialized using a mesh generated by method 2. The initial grid for method 1 or method 3 corresponds to the row on the left hand side of the table, which indicates the grid spacing (uniform grid) or number of points used (method 2), e.g. $n2$ is two points.

We will first discuss the initial meshes: the grid mesh and method 2. For system A, no viable Lyapunov function was found using the grid meshes considered. A grid mesh was able to produce a Lyapunov function using $624$ (system B) and $512$ (system C) simplices. For system A, method 2 was similarly not able to produce a Lyapunov function with the meshes considered, but for systems B and C, it performed worse than a grid mesh -- requiring $1452$ and $1224$ simplices, respectively, to produce a viable Lyapunov function.

The adaptive meshing methods, method 1 (uniform grid initialization + adaptive meshing) and method 3 (method 2 initialization + adaptive meshing), both consistently outperformed a uniform grid on all three dynamical systems. Method 1 found a viable Lyapunov function with 190 simplices (system A), 210 simplices (system B), and 186 simplices (system C). Method 3 found a viable Lyapunov function with 199 simplices (system A), 255 simplices (system B), and 334 simplices (system C). Direct comparison of method 1 and method 3 is difficult, as the two start with different initializations. We consider $\Delta m_\Tcal$, the number of simplices added to the initial mesh before a viable Lyapunov function was found. With this metric, method 1 outperformed method 3 on system A (method 1 requiring 28 additional simplices at its best, while method 3 required 140). Method 3 outperformed method 1 for system C, requiring 2 or 4 additional simplices, where method 1 required 58 at its best. Finally, method 1 and 3 have similar performance on system B -- needing 4 and 8 additional simplices, respectively.

\begin{table}
    \centering
    \begin{subtable}{\columnwidth}\renewcommand{\arraystretch}{1.2}
        \centering
        \begin{tabular}{|C{0.65cm}|C{0.35cm}|C{0.4cm}|C{0.6cm}||C{0.15cm}|C{0.35cm}|C{0.4cm}|C{0.6cm}|C{0.6cm}|}
		\hline
        \rowcolor{LightCyan}
		\multicolumn{4}{|c|}{Grid Mesh} & \multicolumn{5}{|c|}{Method 1} \\
		\hline
        Grid & N & $m_\Tcal$ & Viable $V$ & It. & N & $m_\Tcal$& Viable $V$ &$\Delta m_\Tcal$ \\ 
        \hline
		$\frac{\pi}{2}$ & 9 & 8 & No & 76 & 109 & 195 & Yes & 187 \\ [2pt]
        \hline
		$\frac{\pi}{4}$ & 25 & 32 & No & 64 & 109 & 196 & Yes & 164 \\ [2pt]
        \hline
		$\frac{\pi}{6}$ & 49 & 72 & No & \textbf{51} & \textbf{108} & \textbf{190} & \textbf{Yes} &\textbf{ 118} \\ [2pt]
        \hline
		$\frac{\pi}{8}$ & 81 & 121 & No & 27 & 116 & 198& Yes & 77 \\ [2pt]
		\hline
		$\frac{\pi}{10}$ & 121 & 200 & No & 18 & 145 & 248& Yes & 48 \\ [2pt]
		\hline
		$\frac{\pi}{12}$ & 169 & 288 & No & 11 & 183 & 316& Yes & 28 \\ [2pt]
		\hline
	\end{tabular}
    \end{subtable}
    \begin{subtable}{\columnwidth}\renewcommand{\arraystretch}{1.2}
        \centering
        
        \begin{tabular}{|C{0.65cm}|C{0.35cm}|C{0.4cm}|C{0.6cm}||C{0.15cm}|C{0.35cm}|C{0.4cm}|C{0.6cm}|C{0.6cm}|}
		\hline
        \rowcolor{LightCyan}
		\multicolumn{4}{|c|}{Method 2} & \multicolumn{5}{|c|}{Method 3} \\
		\hline
		Grid & N & $m_\Tcal$ & Viable $V$ & It. & N & $m_\Tcal$ & Viable $V$ & $\Delta m_\Tcal$\\ 
        \hline	
        n2 & 35 & 48 & No & \textbf{54} & \textbf{112} & \textbf{199} & \textbf{Yes} & \textbf{151}\\ 
        \hline
		n3  & 42 & 60 & No & 63 & 125 & 224 & Yes& 164\\
        \hline
		  n4& 49 & 72 & No & 54 & 119 & 212 & Yes & 140\\ 
		\hline
		  n5 & 56 & 84 & No & 54 & 128 & 227 & Yes & 143\\
		\hline
	\end{tabular}
    \caption{System A: Pendulum}
    \label{tab:sysA}
    \end{subtable}
    
    \begin{subtable}{\columnwidth}
            \centering    
       \begin{tabular}{|C{0.65cm}|C{0.35cm}|C{0.4cm}|C{0.6cm}||C{0.15cm}|C{0.35cm}|C{0.4cm}|C{0.6cm}|C{0.6cm}|}
		\hline
        \rowcolor{LightCyan}
		\multicolumn{4}{|c|}{Grid Mesh} & \multicolumn{5}{|c|}{Method 1} \\
		\hline
		  Grid & N & $m_\Tcal$ & Viable $V$ & It. & N & $m_\Tcal$ & Viable $V$  &$\Delta m_\Tcal$\\ 
        \hline	
        0.375 & 25 & 32 & No & \textbf{72} & \textbf{125} & \textbf{210 }& \textbf{Yes}& \textbf{178} \\ 
        \hline
        0.25 & 49 & 72 & No & 65 & 132 & 224 & Yes & 152 \\ 
        \hline
		0.125 & 169 & 288 & No & 3 & 171& 292& Yes & 4\\
        \hline
		  \textbf{0.0625} & \textbf{625} & \textbf{1152}& \textbf{Yes} & --& --&-- & --& -- \\ 
		\hline
		\hline	
	\end{tabular}
    \end{subtable}

    \begin{subtable}{\columnwidth}
            \centering
        
         \begin{tabular}{|C{0.65cm}|C{0.35cm}|C{0.4cm}|C{0.6cm}||C{0.15cm}|C{0.35cm}|C{0.4cm}|C{0.6cm}|C{0.6cm}|}
		\hline
        \rowcolor{LightCyan}
		\multicolumn{4}{|c|}{Method 2} & \multicolumn{5}{|c|}{Method 3} \\
		\hline
		Grid & N & $m_\Tcal$ & Viable $V$ & It. & N & $m_\Tcal$ & Viable $V$ &$\Delta m_\Tcal$ \\ 
        \hline	
        n2 & 70 & 108 & No &76 &172 &300 &Yes & 192\\ 
        \hline
		n3 &117 & 192& No & \textbf{41}&\textbf{164} &\textbf{279} &\textbf{Yes }& 87\\
        \hline
		  n4 &176 &300 & No&33 & 221& 384& Yes & 84\\ 
		\hline
		  n5 & 247 & 432&No &25 &281 &498 &Yes & 66\\
		\hline
		  n6 &330 &588 &No &26 &364 & 654& Yes& 66\\ 
		\hline	
        n7 & 452 & 768& No& 19& 443& 802& Yes& 34\\ 
		\hline	
        n8 & 532 & 972& No& 25&551 &1007 &Yes & 35\\ 
		\hline	
        n9 & 651 &1200 &No &5 & 655& 1208&Yes & 8\\ 
		\hline	
        \textbf{n10} & \textbf{782} & \textbf{1452}& \textbf{Yes}& --& --& --& --& --\\ 
		\hline	
	\end{tabular}
    \caption{System B}
    \label{tab:sysB}
    \end{subtable}

    \begin{subtable}{\columnwidth}
            \centering    
        \begin{tabular}{|C{0.65cm}|C{0.35cm}|C{0.4cm}|C{0.6cm}||C{0.15cm}|C{0.35cm}|C{0.4cm}|C{0.6cm}|C{0.6cm}|}
		\hline
        \rowcolor{LightCyan}
		\multicolumn{4}{|c|}{Grid Mesh} & \multicolumn{5}{|c|}{Method 1} \\
		\hline
		Grid & N & $m_\Tcal$ & Viable $V$ & It. & N & $m_\Tcal$ & Viable $V$ & $\Delta m_\Tcal$ \\ 
        \hline	
         0.5 & 25 & 32 & No & 76 & 120 & 206 & Yes&174\\ 
        \hline
		  0.$\bar{3}$ & 49 & 72 & No &55 &118 &200 &Yes&128\\
        \hline
		0.25 & 81& 128& No& \textbf{28}& \textbf{110}&\textbf{186} &\textbf{Yes}& \textbf{58}\\ 
		\hline
		  \textbf{0.125} & \textbf{289}& \textbf{512}&\textbf{Yes} & --&--&-- & --& --\\
		\hline
	\end{tabular}
    \end{subtable}

    \begin{subtable}{\columnwidth}
            \centering

         \begin{tabular}{|C{0.65cm}|C{0.35cm}|C{0.4cm}|C{0.6cm}||C{0.15cm}|C{0.35cm}|C{0.4cm}|C{0.6cm}|C{0.6cm}|}
		\hline
		\multicolumn{4}{|c|}{Method 2} & \multicolumn{5}{|c|}{Method 3} \\
		\hline
        \rowcolor{LightCyan}
		Grid & N & $m_\Tcal$ & Viable $V$ & It. & N & $m_\Tcal$ & Viable $V$ & $\Delta m_\Tcal$ \\ 
        \hline	
        n2 & 77 &120 & No & 86 & 211 & 380 & Yes &260\\ 
        \hline
		n3 & 135 & 224 & No &\textbf{43} & \textbf{191}& \textbf{334}&\textbf{Yes} &\textbf{110}\\
        \hline
		  n4 & 209& 360& No&13 & 228& 398&Yes&38\\ 
		\hline
		  n5 &299 & 528& No &3 & 301& 532& Yes&4\\
		\hline
		  n6 &405 & 728& No& 3&407 &732 & Yes&4\\ 
        \hline
        n7 &527 & 960& No& 2& 528&962 & Yes&2\\ 
        \hline
        \textbf{n8} &\textbf{665} &\textbf{1224} & \textbf{Yes} & --& --&-- &-- &\\ 
		\hline	
	\end{tabular}
    
    \caption{System C}
    \label{tab:sysC}
    \end{subtable}
    
    \caption{Results of using methods 1 to 3 to generate meshes over $\Omega$ to produce \ac{cpa} Lyapunov functions for each two-dimensional system. Here,  $N$ indicates the number of mesh vertices, and $m_\Tcal$ denotes the number of simplices. For methods 1 and 3, the number of iterations (It) and total number of simplices added to the initial mesh ($\Delta m_\Tcal$) are recorded. The best result of each method is in bold. }
    \label{tab:results2D}
\end{table}

\begin{figure}[h]
    \centering
    \begin{subfigure}[t]{0.5\columnwidth}
        \centering
        \includegraphics[width= \textwidth]{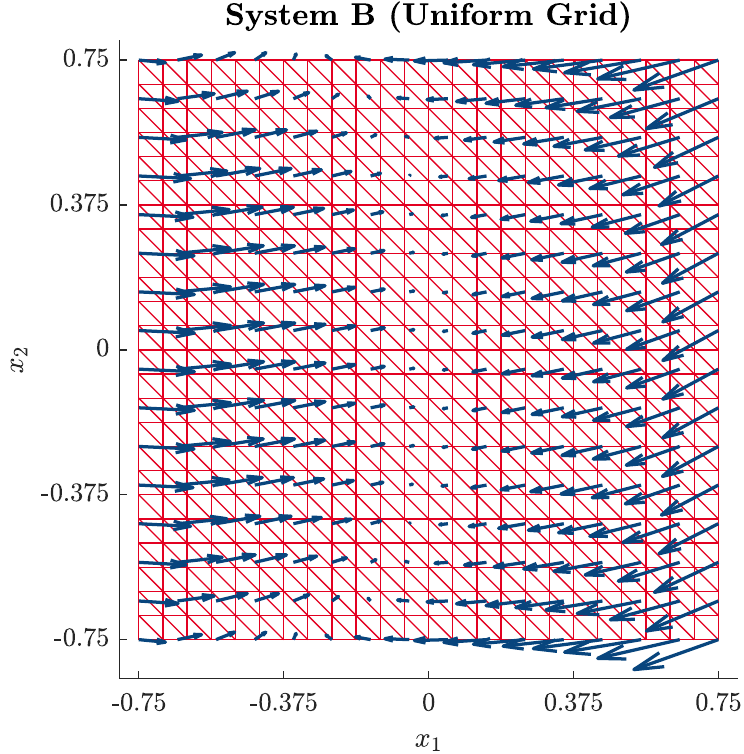}
        \caption{}
        \label{fig:sysB_grid}
    \end{subfigure}%
    \hfill
    \begin{subfigure}[t]{0.5\columnwidth}
        \centering
        \includegraphics[width = \textwidth]{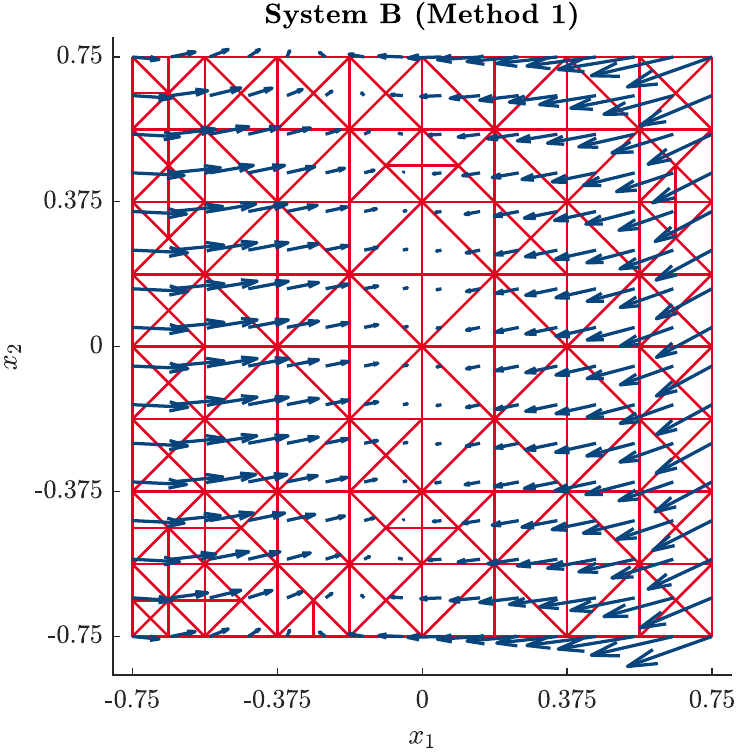}
        \caption{}
        \label{fig:sysB_m1}
    \end{subfigure}

    \centering
    \begin{subfigure}[t]{0.5\columnwidth}
        \centering
        \includegraphics[width= \textwidth]{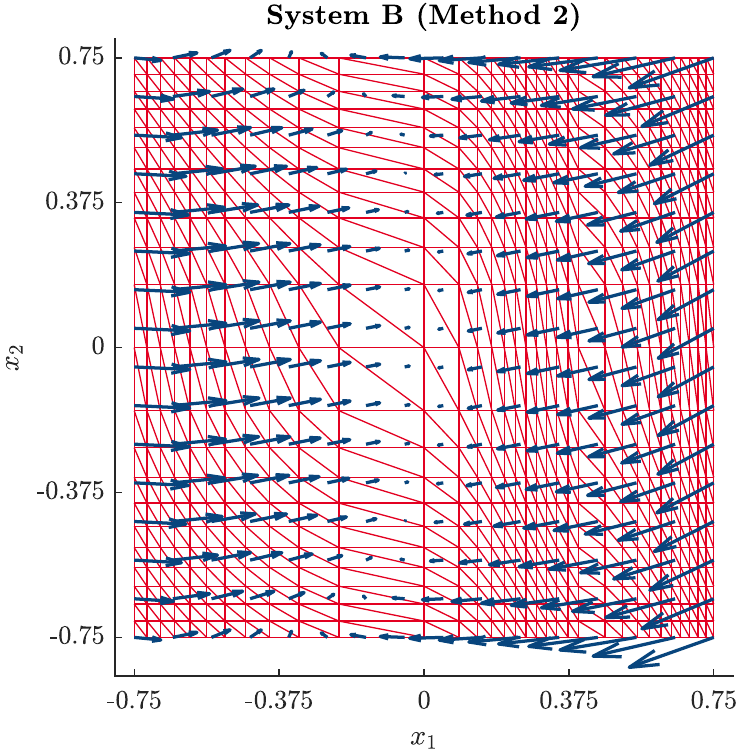}
        \caption{}
        \label{fig:sysB_m2}
    \end{subfigure}%
    \hfill
    \begin{subfigure}[t]{0.5\columnwidth}
        \centering
        \includegraphics[width =\textwidth]{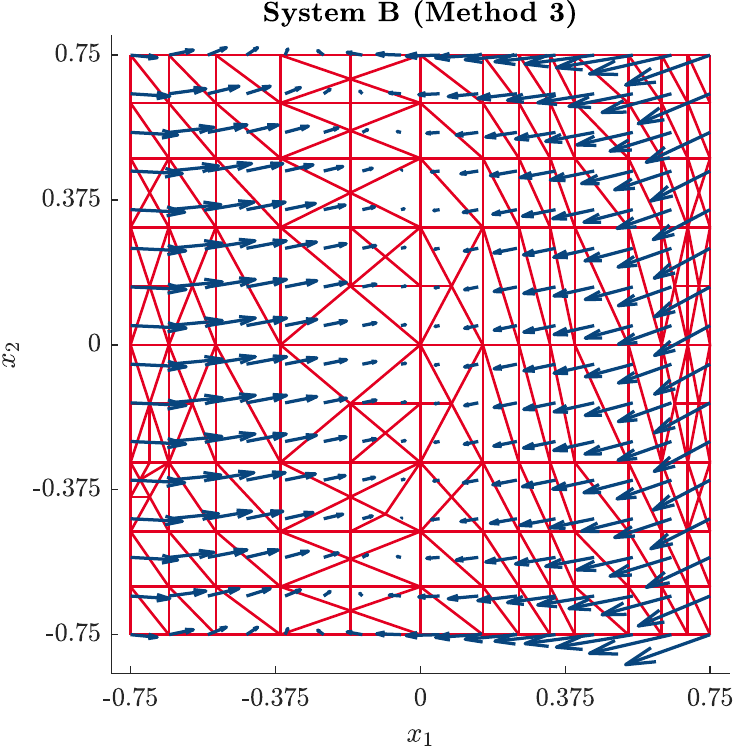}
        \caption{}
        \label{fig:sysB_m3}
    \end{subfigure}
    \caption{Comparison of the best meshes (meshes with the lowest number of simplices able to produce a viable Lyapunov function) for system B plotted over the phase plot of the system with arrow scale 1.75. Here, \ref{fig:sysB_grid} is the grid mesh that found a valid Lyapunov function using the fewest simplices, while \ref{fig:sysB_m1} shows method 1, \ref{fig:sysB_m2} shows method 2, and \ref{fig:sysB_m3} shows method 3.}
    \label{fig:sysB}
\end{figure}

\begin{figure}[h]
    \centering
    \begin{subfigure}[t]{0.5\columnwidth}
        \centering
        \includegraphics[width= \textwidth]{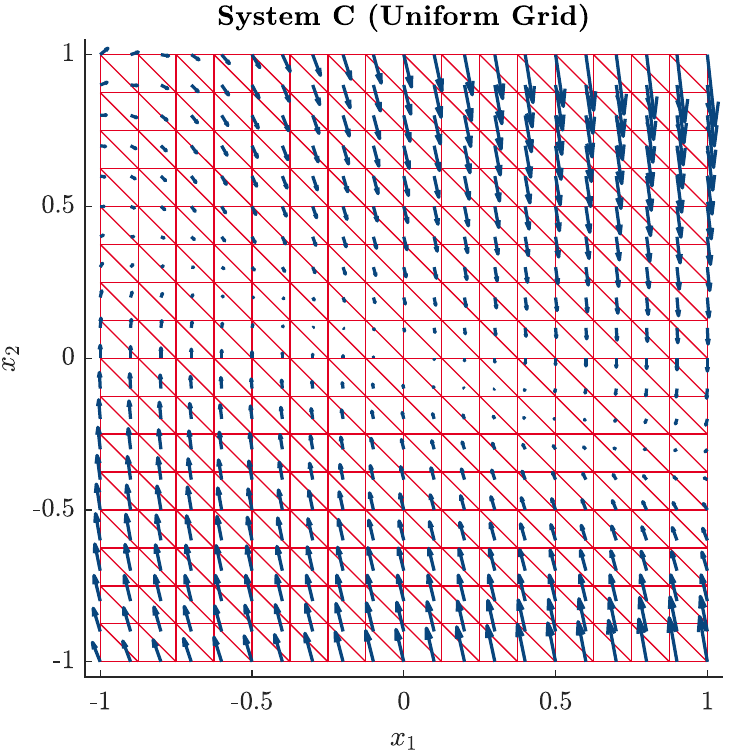}
        \caption{}
        \label{fig:sysC_grid}
    \end{subfigure}%
    \hfill
    \begin{subfigure}[t]{0.5\columnwidth}
        \centering
        \includegraphics[width=\textwidth]{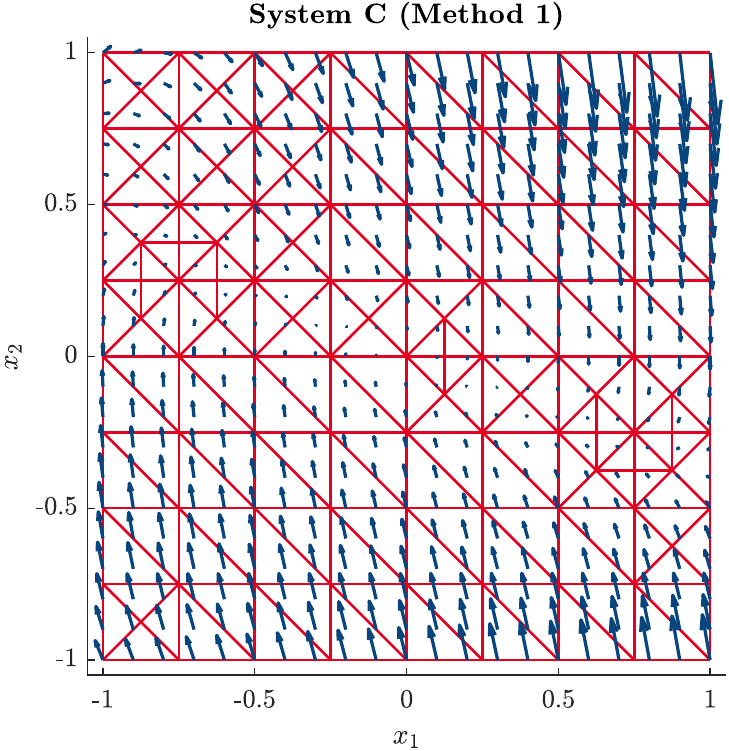}
        \caption{}
        \label{fig:sysC_m1}
    \end{subfigure}

    \centering
    \begin{subfigure}[t]{0.5\columnwidth}
        \centering
        \includegraphics[width= \textwidth]{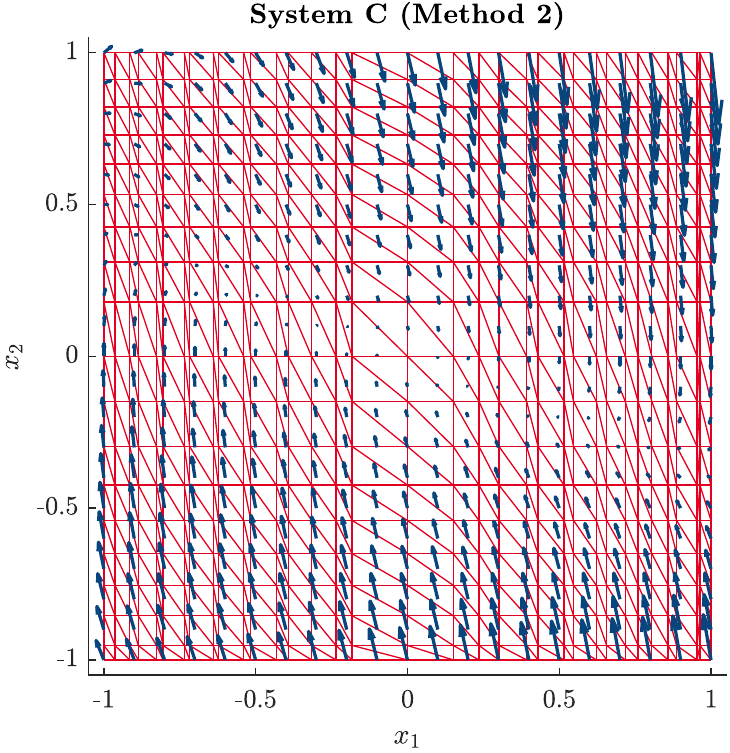}
        \caption{}
        \label{fig:sysC_m2}
    \end{subfigure}%
    \hfill
    \begin{subfigure}[t]{0.5\columnwidth}
        \centering
        \includegraphics[width = \textwidth]{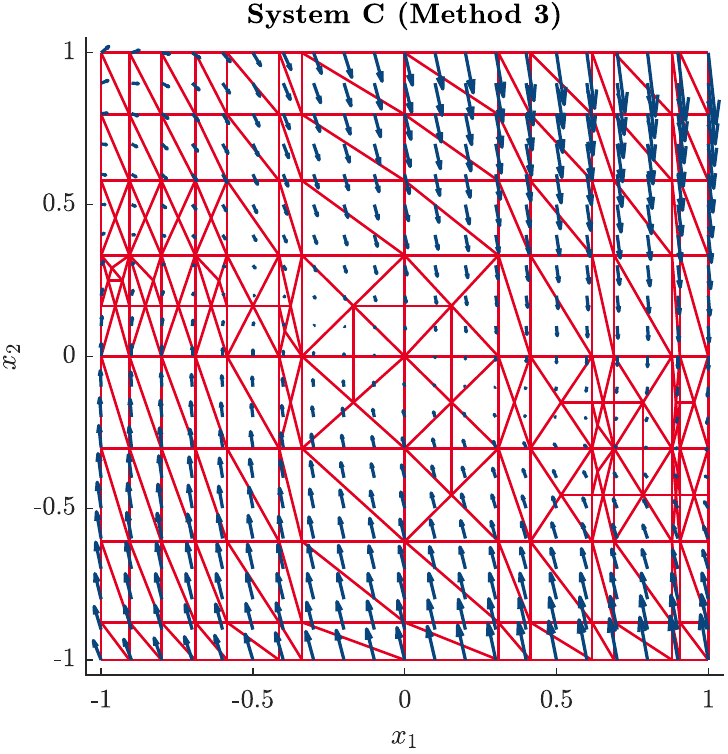}
        \caption{}
        \label{fig:sysC_m3}
    \end{subfigure}
    \caption{Comparison of the best meshes (meshes with the lowest number of simplices able to produce a viable Lyapunov function) for system C plotted over the phase plot of the system with arrow scale 1.75. Here, \ref{fig:sysC_grid} is the grid mesh that found a valid Lyapunov function using the fewest simplices, while \ref{fig:sysC_m1} shows method 1, \ref{fig:sysC_m2} shows method 2, and \ref{fig:sysC_m3} shows method 3.}
    \label{fig:sysC}
\end{figure}

Figures \ref{fig:sysB} and \ref{fig:sysC} show the best performing mesh (i.e., mesh with the lowest number simplices able to produce a viable Lyapunov function) of each method for system B and system C, respectively, as well as the phase plot of the system. The adaptive meshing methods (method 1 and method 3) often refined areas where the phase plane was near zero, while method 2 focused a on areas where the most change in curvature of the system dynamics is occurring.

\subsection{Three Dimensional Nonlinear System}

We also considered the three dimensional system,
\begin{flalign}\label{eq:sysD}
    \dot{\x} = \begin{pmatrix}
            f_1(\x) \\
		     f_2(\x) \\
             f_3(\x)
        \end{pmatrix} =
            \begin{pmatrix}
    	       -3x_1 + 0.5x_1-x_3x_2^4\\
    		   -x_2x_3^4-2.5x_2 + 0.5x_3 \\
               -0.5x_2 - 5x_3 + x_1x_2^2,
    	\end{pmatrix}
\end{flalign}
over $\Omega = [-1,1]\times [-1, 1]\times [-1,1].$ The Hessian of $f_1(\x)$ contains the terms $-4x_2^3$ and $-12x_3x_2^2$, the Hessian of $f_2(\x)$ depends on the values of $-4x_3^3$ and $-12x_3^2x_2$, and the Hessian of $f_3(\x)$ depends on $2x_1$ and $2x_2.$ Because $x_1$ is linear in the Hessian, method 2 does not produce bounds for this dimension. We use a uniform gridding of $0.25$ for $x_1.$ 

\Cref{tab:sysD} shows the results of applying methods 1-3 when compared to a uniform grid mesh. Here, the importance of exploring different meshing strategies becomes apparent, as the three dimensional system requires $24,576$ simplices to compute a viable Lyapunov function with a uniform grid meshing. In contrast, methods 1 and 3 find viable Lyapunov functions with $3,431$ and $4,390$ simplices, respectively.

\begin{table}
    \centering
    \begin{subtable}{\columnwidth}
            \centering
         
      \begin{tabular}{|C{0.65cm}|C{0.4cm}|C{0.55cm}|C{0.6cm}||C{0.25cm}|C{0.375cm}|C{0.4cm}|C{0.6cm}|C{0.6cm}|}
		\hline
        \rowcolor{LightCyan}
		\multicolumn{4}{|c|}{Grid Mesh} & \multicolumn{5}{|c|}{Method 1} \\
		\hline
		Grid  & N & $m_\Tcal$ & Viable $V$ & It & N & $m_\Tcal$ & Viable $V$ & $\Delta m_\Tcal$ \\ 
        \hline	
         1 &27 &28 &No  & 527& 878 & 3940 & Yes &3912\\ 
        \hline	
          0.5&125 &384 &No  & 485& 899 & 4009 & Yes &3625\\ 
        \hline
		  0.25 & 729 & 3072& No&\textbf{53} & \textbf{809}&\textbf{3431} &\textbf{Yes} &\textbf{359} \\
        \hline
		  \textbf{0.125} & \textbf{4913}& \textbf{24576} &\textbf{Yes} & -- &  --& --&-- &--\\ 
		\hline
	\end{tabular}
    
    \end{subtable}
    
    \begin{subtable}{\columnwidth}
            \centering
        
        \begin{tabular}{|C{0.65cm}|C{0.4cm}|C{0.55cm}|C{0.6cm}||C{0.25cm}|C{0.375cm}|C{0.4cm}|C{0.6cm}|C{0.6cm}|}
		\hline
        \rowcolor{LightCyan}
		\multicolumn{4}{|c|}{Method 2} & \multicolumn{5}{|c|}{Method 3} \\
		\hline
		Grid & N & $m_\Tcal$ & Viable $V$ & It. & N & $m_\Tcal$ & Viable $V$&$\Delta m_\Tcal$\\ 
        \hline
		  n3 & 891 &3840 &No &\textbf{64} &\textbf{1009}  & \textbf{4390} & \textbf{Yes} &\textbf{550} \\
        \hline
		  n4& 1485& 6720& No & 10 &1505& 6820 &Yes &100\\ 
        \hline	
        n5 &1989 & 9216 & No&10 &2013 &9336 & Yes&120\\ 
		\hline
	\end{tabular}
    
    \caption{System 3b}
    \end{subtable}%
    \caption{Results of using methods 1-3 to generate meshes over $\Omega$ for \ac{cpa} Lyapunov function synthesis for \eqref{eq:sysD}. The number of vertices ($N$) and simplices ($m_\Tcal$) of each mesh are indicated. The uniform grid spacing or number of bounding points (method 2) are indicated in the left most column. For methods 1 and 3, the number of iterations (It) and the number of simplices added to the initial mesh ($\Delta m_\Tcal$) are recorded. The best result of each method is in bold.}
    \label{tab:sysD}
\end{table}

\section{DISCUSSION}

Numerical experiments showed that both adaptive meshing methods (method 1 and method 3) performed better than a uniform grid mesh regardless of initialization. For the three dimensional case, the number of simplices was an order of magnitude lower than that of the uniform grid mesh. %When comparing method 1 and method 3, method 1 seemed to have better results over all. In Figures \ref{fig:sysB_m1} and \ref{fig:sysC_m1}, we see that 
Method 1 (\ref{fig:sysB_m1} and \ref{fig:sysC_m1}) creates more regular meshes, while method 3 (\ref{fig:sysB_m3} and \ref{fig:sysB_m3}) often produces irregular shapes due to the initial mesh. Method 3 often required less additional simplices when adaptively refining, which may be a result of the initial grid beginning with more simplices. 
%method 2 initializing with more simplices to start. %There is a trade off in computational complexity in that method 1 and method 3 both require iteratively solving an optimization problem. %This trade off is affected by system dimension. The \ac{lp} is solved via interior point methods, which have polynomial time complexity \cite{nemirovski2004interior}. \blue{ Thus, as the system's state dimension increases, it preferable to iteratively solve a smaller problem. }

Method 2 performed the same or worse than a uniform grid mesh. Because Method~2 constructs candidate points by examining the univariate decomposition of system dynamics, it may yield limited insight when the univariate convexity structure is locally constant or when the Hessian is dominated by off-diagonal terms.
%Method~2 constructs candidate points by decomposing the system dynamics into their univariate components and generating samples at locations where the convexity of each univariate term changes. 
%Although this approach is effective for deriving tight bounds, it may yield limited insight when the univariate convexity structure is locally constant or when the Hessian is dominated by off-diagonal terms, causing multivariate curvature effects that are not captured by the univariate decomposition.
It is expected that method 2 might perform poorly on system C (due to the dominance of off diagonal terms in the Hessian). 
Further, the structure of system~A is relatively simple within $\Omega$, yielding limited insight from method~2. However, it is surprising that method~2 did not perform better on system~B. This leads to a key insight.% in these results.
%Further, system A, which has a sinusoidal term in the Hessian, is likely not dynamic enough to prompt insight from method 2. 

%This is not an unexpected result. \red{Method 2 produces points that allow for the system dynamics to be bounded tightly}. For system A, the singular sinusoidal term in the Hessian is likely not dynamic enough to prompt insight from method 2. \red{Systems B and C are more dynamic. However,} because method 2 considers univariate functions, it is unable to fully consider how interactions between system variables affect the convexity or concavity of the system dynamics. This biases the algorithm towards nonlinear terms that are univariate (system B).

Our results show that the region of the state space that requires a finer mesh is not always the region with large or changing Hessian terms, but is often the region where $f(\x)$ is near zero. In Figures \ref{fig:sysC_m1} and \ref{fig:sysC_m3} (system C), adaptive mesh refinement targets where $f(\x)$ is near 0, likely because satisfying the Lyapunov decrease condition is more difficult as $f(\x) \rightarrow 0.$ The error term in \eqref{eq:decrease} dominates the inequality and must be reduced before \eqref{eq:decrease} can be satisfied. In Figures \ref{fig:sysB_m1} and \ref{fig:sysB_m3} (system B), this effect is less pronounced, as refinement is also targeted in regions where $f(\x)$ is larger, but still visible. Overall, these results show that meshing should consider the error term relative to the decrease condition.

\section{CONCLUSION}
This paper examines different meshing strategies to more efficiently synthesize \ac{cpa} Lyapunov functions for nonlinear dynamical systems. Numerical results demonstrate that adaptive meshing guided by an \ac{lp} outperforms a naive uniform grid mesh. These results also shed light on potential future mesh initializations that prioritize dense meshing in areas with little system evolution, i.e., where $f(\x)$ is near $0$.

While adaptive meshing provides a promising direction, its requirement of iteratively re-running an \ac{lp} is computationally expensive. Future work aims to establish if singular simplices or groups of simplices can be refined without changing the solution of the Lyapunov function elsewhere -- maintaining the benefits of method 1, while reducing computational load. %Future work should also evaluate mesh refinement methods on dynamical systems with higher state dimensions. %More initialization schemes for meshing should be explored in the future as well. %For example,  method 2 should be improved to consider interactions between system variables in determining the convexity of a multivariate function to provide more salient initial meshes. 
%For example, exploring meshes that prioritizes density of simplices in areas where $f(\x)$ is near $0$ is an interesting direction.
%, in light of the results from systems B and C.

\addtolength{\textheight}{-12cm}   % This command serves to balance the column lengths

\bibliographystyle{IEEEtran}
\bibliography{IEEEabrv,biblio}
\end{document}

%% file: Packages.tex
\usepackage{amsmath,amssymb,bm} % for math symbols and notation
\usepackage{acronym}
\usepackage{color}
\usepackage{hyperref}
\usepackage{graphicx}		
\usepackage{mathtools}
\usepackage[normalem]{ulem}
\usepackage{comment}
\usepackage{tikz}
\usepackage{cleveref}
\usepackage{algorithm,algpseudocode}
\usepackage{multicol}
\usepackage{multirow}
\usepackage{subcaption}
\usepackage{array}
\usepackage{pdfpages}
\usepackage{pgfplots}
\usetikzlibrary{shapes.geometric}
\usetikzlibrary{arrows.meta,arrows}
\usepackage{xcolor,colortbl}

%% file: CustomCommands.tex
\definecolor{mygray}{gray}{0.35}
\newcommand{\red}[1]{\textcolor{red}{#1}} % red writing
\definecolor{LightCyan}{rgb}{0.88,1,1}

\def\IR{\mathbb R}					% Real #'s
\newtheorem{definition}{Definition}[section]
\newtheorem{theorem}{Theorem}
\newtheorem{lemma}{Lemma}
\newtheorem{corollary}{Corollary}

\newtheorem{statement}{Objective}

\newcommand{\mbf}[1]{\mathbf{#1}} % bold
\newcommand{\x}{\mathbf{x}} % bold x
\newcommand{\W}{\mbf{W}}

\newcommand{\X}{\mbf{X}}

\newcommand{\Tcal}{\mathcal{T}}

\newcommand{\vecdim}[1]{\in \mathbb{R}^{#1}}

\newcommand{\norm}[1]{\left\lVert#1\right\rVert}
\newcommand{\absVal}[1]{\left\lvert#1\right\rvert}

\usepackage{centernot}
\acrodef{dl}[{DL}]{deep learning}
\acrodef{rl}[{RL}]{reinforcement learning}
\acrodef{nn}[{NN}]{neural network}
\acrodef{dnn}[{DNN}]{deep neural network}
\acrodef{tdl}[{TDL}]{temporal difference learning}
\acrodef{pid}[{PID}]{proportional–integral–derivative}
\acrodef{us}[{US}]{Ultrasound}
\acrodef{mse}[{MSE}]{mean squared error}
\acrodef{sgd}[{SGD}]{stochastic gradient descent}
\acrodef{ico}[{ICO}]{iterative convex overbounding}
\acrodef{lmi}[{LMI}]{linear matrix inequality}
\acrodef{mjls}[{MJLS}]{Markov jump linear system}
\acrodef{io}[{IO}]{input-output}
\acrodef{iqc}[{IQC}]{integral quadratic constraints}
\acrodef{cnn}[{CNN}]{convolutional neural network}
\acrodef{il}[{IL}]{Imitation learning}
\acrodef{mpc}[{MPC}]{model predictive control}
\acrodef{sdp}[{SDP}]{semi-definite programming}
\acrodef{relu}[{ReLU}]{rectified linear unit}
\acrodef{us}[US]{ultrasound}
\acrodef{mdp}[MDP]{Markov Decision Process}
\acrodef{iid}[iid]{identical and independently distributed random variable}
\acrodef{pid}[PID]{Proportional Integral Derivative}

\acrodef{lqr}[LQR]{linear-quadratic regulator}

\acrodef{cpa}[CPA]{continuous piecewise affine}
\acrodef{hji}[HJI]{Hamilton Jacobi Inequality}
\acrodefplural{hji}[HJI]{Hamilton Jacobi Inequalities}
\acrodef{roa}[ROA]{region of attraction}
\acrodef{lp}[LP]{linear program}
\acrodef{fem}[FEM]{finite element method}
\acrodef{cfd}[CFD]{computational fluid dynamics}

\acrodef{pde}[PDE]{partial differential equation}
\acrodef{ode}[ODE]{ordinary differential equation}
\acrodef{pdi}[PDI]{partial differential inequality}
\acrodef{leb}[LEB]{longest edge bisection}

\newcommand{\PreserveBackslash}[1]{\let\temp=\\#1\let\\=\temp}
\newcolumntype{C}[1]{>{\PreserveBackslash\centering}p{#1}}
\newcolumntype{P}[1]{>{\centering\arraybackslash}p{#1}}